\documentclass[10pt]{article} 
\usepackage{aas2pp4}

\renewcommand{\H}{\mbox{\it H}}
\newcommand{\K}{\mbox{\it K}}

\newcommand{\AK}{\mbox{$A_{\it K}$}}

\newcommand{\mk}{\mbox{$m_{\it K}$}}
\newcommand{\mbol}{\mbox{$m_{\rm bol}$}}

\newcommand{\Msun}{\mbox{$M_\odot$}}

\newcommand{\HI}{\ion{H}{1}}
\newcommand{\HII}{\ion{H}{2}}
\newcommand{\HeI}{\ion{He}{1}}
\newcommand{\HeII}{\ion{He}{2}}
\newcommand{\NIII}{\ion{N}{3}}
\newcommand{\CIV}{\ion{C}{4}}

\newcommand{\UCHII}{UC~\HII}

\renewcommand{\[}{\begin{eqnarray}}
\renewcommand{\]}{\end{eqnarray}}
 
\newcommand{\Br}{{\rm Br\ifmmode \gamma \else $\gamma$\fi}}

\newcommand{\g}{G29.96$-$0.02}

\begin{document}

\title{
Near Infrared Spectroscopy of {\g}:\\
The First Spectral Classification of the Ionizing Star\\
of an Ultracompact H~{\large II} Region
}

\author{
Alan M. Watson\footnote{Current address: Instituto de Astronom{\'\i}a, UNAM,
J.~J.~Tablada 1006, Col.\ Lomas de Santa Maria, 58090 Morelia, Michoac{\'a}n,
Mexico; alan@astrosmo.unam.mx}
}
\affil{
Department of Astronomy,
New Mexico State University,
Las Cruces,
NM 88001
}
\author{
Margaret M. Hanson\footnote{Hubble Fellow}
}
\affil{
Steward Observatory,
University of Arizona,
Tucson,
AZ 85721\\
mhanson@as.arizona.edu
}

\begin{abstract}
\leftskip=1in
\rightskip=\leftskip
\normalsize
\noindent
We have obtained the first classification spectrum and present the first
direct spectral classification of the ionizing star of an ultracompact
{\HII} region. The ultracompact {\HII} region is {\g}, a well-studied
object with roughly twice solar metallicity. The near infrared {\K}-band
spectrum of the ionizing star exhibits {\CIV} and {\NIII} emission and
{\HeII} absorption, but lines of {\HI} and {\HeI} are obliterated by
nebular emission. We determine that the star has a spectral type of O5
to O7 or possibly O8.  We critically evaluate limits on the properties
of the star and find that it is compatible with zero-age main-sequence
properties only if it is binary and if a significant fraction of the
bolometric luminosity can escape from the region. {\g} will now be an
excellent test case for nebular models, as the properties of the ionizing
star are independently constrained.

\end{abstract}

\begin{center}
28 August 1997\\
Accepted for publication in ApJ Letters
\end{center} 

%ms: \keywords{
%ms: infrared: stars -- 
%ms: stars: early-type -- 
%ms: ISM: {\HII} regions -- 
%ms: ISM: individual {\g}
%ms: }

\section{Introduction}

Ultracompact (UC) {\HII} regions are formed by massive stars still
embedded in their natal molecular clouds. They are luminous and
relatively easy to detect throughout the galaxy at far infrared and
centimeter wavelengths (Wood \& Churchwell 1989). As such, they hold
great potential for the study of the formation and properties of massive
stars in different environments and at different metallicities. However,
the many magnitudes of extinction to these objects has hindered their
study; until recently the stars have had to be studied by their indirect
and uncertain effects on the interstellar medium.

In this {\it Letter} we present the first near infrared classification
spectrum of the ionizing star of an {\UCHII} region and directly derive
narrow limits on its effective temperature by comparison to the stellar
atlas of Hanson, Conti, \& Rieke (1996). The {\UCHII} region 
is {\g}, a well-studied object with a metallicity of about twice solar
(Simpson et~al.\ 1995; Afflerbach, Churchwell, \& Werner 1997).  Our work
is complementary to that of Watson et~al.\ (1997), who directly measured
the extinction and intrinsic magnitude of the star using imaging in
the radio and near infrared and simple nebular physics. Combining these
results, we are able to place the star on the H-R diagram with a high
degree of reliability.

Our work motivates a critical re-evaluation of the conclusion of Watson
et~al.\ (1997), that the ionizing star of {\g} shows evidence for the
evolution predicted by Bernasconi \& Maeder (1996), and our independent
constraints on the properties of the ionizing star will provide an
important test for nebular models.  However, our main result is to
demonstrate that the ionizing stars of at least some {\UCHII} regions
are now open for business in the near infrared.

\section{Data}
\label{sec-data}

We obtained spectra of {\g} on the night of 1997 June 18 UT with the
FSPEC spectrograph (Williams et al.\ 1993) on the Multiple Mirror
Telescope. The seeing was about 1 arcsec FWHM and the sky was clear.
The spectrograph has a $1.2 \times 32$ arcsec slit and gave coverage
from about 2.03{\micron} to 2.21{\micron} at a resolution of
$\lambda/\Delta\lambda \approx 1000$.

We obtained twelve 4 minute spectra of the ionizing star of {\g} to give
a total 48 minutes of exposure.  We interspersed observations of {\g}
with observations of the nearby bright A1V star HR~7209.

Figure~\ref{fig-image} shows near infrared {\Br} and {\K}-band images
of {\g} taken from Watson et~al.\ (1997). The images show clearly the
ionizing star at the center of a bright arc of nebular emission. A
fainter fan of emission extends away from the arc to the north east.
The spectrograph incorporates a slit viewing camera working in the
{\H}-band and so we were able to accurately position the slit and guide
manually.  We oriented the slit in elevation to simplify guiding and as
a result the slit rotated on the sky between position angles of 6~deg
and 40~deg.  We nodded the ionizing star along the slit.

We extracted the stellar spectrum in a synthetic aperture of 1.2 arcsec
width centered on the ionizing star. This aperture matches both the image
quality and the slit width. We also extracted spectra of the bright arc
of the nebula to the south and west and the fainter, extended fan of
the nebula to the north and east using apertures from 1.2 to 3.6 arcsec
either side of the ionizing star.

We corrected for atmospheric absorption using the spectra of HR~7209. We
removed the {\Br} absorption by fitting and adding a Gaussian.
While this will leave systematic errors of as much as 10\% in this
region, the {\Br} line is of limited use as it is heavily
contaminated by nebular emission. We used a black-body with a
temperature of 9970~K (Code et al.\ 1976) for the intrinsic continuum
of HR~7209.

Our twelve individual spectra of {\g} exhibited subtle ripples over
scales of tens of pixels, the result, we think, of our relatively narrow
synthetic aperture. To counter these ripples, we divided each spectrum
by a fit to the continuum, then averaged the ratios (with 2$\sigma$
rejection) and the fits separately, and finally multiplied the average
ratio by the average continuum fit to give an average spectrum. The
distribution of the individual spectra about the average spectrum implies
a S/N of about 300 and examination of regions of the spectrum away from
nebular, photospheric, and teluric features suggests that this S/N is
indeed being achieved, at least in some places.

Figure~\ref{fig-flux}a shows the nebular arc spectrum. The fan
spectrum is similar, except that it has a lower S/N and the {\HeI}
2.1120{\micron} and 2.1132{\micron} lines are relatively weaker by
about 10\%. Figure~\ref{fig-flux}b shows spectra of the ionizing star.
The spectrum marked `A' is the total spectrum with no attempt to
remove the nebular contamination. To create the spectrum marked `B',
we scaled and subtracted the arc nebular spectrum so that the
{\Br} and {\HeI} 2.0581{\micron} lines were neither in emission
or absorption. To create the spectrum marked `C', we scaled and
subtracted the arc nebular spectrum so that the {\Br} and {\HeI}
2.0581{\micron} lines were unreasonably deep; the equivalent width of
{\Br} is about 33{\AA} in spectrum C, but in OB stars it is never
deeper than about 10{\AA} (Hanson, Conti, \& Rieke 1996). Thus, the
real spectrum of the ionizing star lies between spectra A and C and is
probably closely approximated by spectrum B, other than near the
strongest nebular lines. (We used the arc spectrum to correct for
nebular emission as it has a higher S/N than the fan spectrum. Thus,
we will likely very slightly over subtract the {\HeI} 2.1120{\micron}
and 2.1132{\micron} lines.) It can be seen from the difference in the
continuum fluxes between spectra A and C in Figure~\ref{fig-flux}b
that the veiling contribution of the nebular continuum is only about
10--20\%.

\section{The Effective Temperature}

The $K$-band spectral features of O stars are almost independent of
luminosity class except for the {\Br} and {\HeI} 2.0581{\micron}
lines, which are obliterated in our stellar spectrum by nebular
emission. Thus, we can determine the effective temperature of the star
from its spectrum but not its luminosity class.

Figure~\ref{fig-spec} shows our normalized spectra along with spectra
from the spectral atlas of Hanson, Conti, \& Rieke (1996) for O stars of
type O3V to O9.5V (HD~93205, HD~168076, HD~93204, Cyg~OB2 516, HD~168075,
HD~467839, HD~101413, and HD~37468). The strong nebular contamination of
the {\Br} and {\HeI} lines means that we have no reliable information
on these lines. Nevertheless, the characteristic spectral features of
other ions are present in the spectrum of the ionizing star. First,
the star clearly has strong {\CIV} emission.  Second, it has {\HeII}
absorption. Third, recalling that spectrum C is oversubtracted and noting
the asymmetry of the stellar 2.11--2.12{\micron} feature compared to the
nebular 2.11--2.12{\micron} feature, the star has {\NIII} emission. The
{\HeII} absorption and {\NIII} emission both imply that the star is kO8
or earlier and the strong {\CIV} emission limits the star to kO5 to kO8
(cf. Table~6 of Hanson, Conti, \& Rieke 1996).  (The k in this notation
is an indication that these are {\K}-band spectral types not optical
MK spectral types.) Normally, the strength of the {\NIII} and {\CIV}
emission features would rule out spectral classes kO7 and kO8 because
these features are weak in solar metallicy stars.  However, the roughly
twice solar metallicity of {\g} may enhance the strength of these
features.

Spectral classes of k05 to kO8 correspond closely to MK spectral classes
of O5 to O8 (Hanson et~al.\ 1996).  This range corresponds to effective
temperatures of about 46000~K to about 38500~K (Vacca, Garmany, \&
Shull 1996).

We also note that the {\CIV} and {\HeII} lines seem broad with FWHM
of about 500~$\rm km\,s^{-1}$; we are planning to obtain spectra at
higher resolution to investigate this further.

\section{The Nature of the Star}

One of the objectives of this study was to check the conclusions of
Watson et~al.\ (1997), who noted that the star had an evolutionary age in
excess of about $10^6$~yr in apparent contradiction to the estimated age
of the {\UCHII} region of only $10^5$~yr. A zero-age main-sequence(ZAMS)
binary could not satisfy their limits and they rejected the possibility
of a close triple system as too unlikely. They suggested that the best
explanation was the idea of Bernasconi \& Maeder (1996), that a massive
star evolves as it accretes.

Figure~\ref{fig-model} shows our limits on the effective temperature
and the limits from Watson et al.\ (1997) on the effective
temperature, luminosity, and distance. 
Also shown are evolutionary tracks and isochrones for the $Z =
2Z_\odot$ models of Meynet et al.\ (1994). Our new limits are
consistent with those of Watson et~al.\ (1997). Taken together, the
limits are consistent with a single or binary star with an apparent age
of between about 1 and $2 \times 10^6$~yr.

Might the apparent age of the {\UCHII} region be wrong? Long lived
{\UCHII} regions have recently been proposed by De~Pree,
Rodr{\'\i}guez, \& Goss (1995) and Garc{\'\i}a-Segura \& Franco
(1996). The basic idea is that the densities in molecular cores can be
large enough to stall the expansion of the {\UCHII} region. We note,
however, that {\g} is not centered on the cloud core (see Figure~6 of
Watson et~al.\ 1997) and appears to have entered a champagne flow
phase (Lumsden \& Hoare 1996). For the {\UCHII} to be of order
$10^6$~yr old would require that its {\HII} region had been confined
for most of this time but recently released. Both the confinement,
given the location of the core, and the fine tuning of the release
seem unlikely to us. We conclude that an age of order $10^5$~yr is
likely to be correct.

Might the limits on the properties of the ionizing star be wrong? The
limits of Watson et~al.\ (1997) on the {\mk} and distance and our
limits on the effective temperature seem to be very robust. The {\mk}
depends only on directly observed quantities and the simple nebular
physics of ionized hydrogen to determine $\AK$. Widening the distance
limit of $5~{\rm kpc} \le d \le 10~{\rm kpc}$ to 4 or 11~kpc would
imply unacceptably large random velocities of $30~\rm km\,s^{-1}$. Our
effective temperature limit is derived by comparison of the observed
spectrum of the ionizing star to well-calibrated spectral standards.

What about the limit on the effective temperature derived by Watson
et~al.\ (1997) from a consideration of the observed {\mbol} and {\mk}?
If, for the moment, we ignore this limit, we can satisfy the other limits
in the small corner of the available parameter space occupied by binaries
of O5 or O5.5 ZAMS stars at 5~kpc. (Earlier or later single stars or
binaries are still forbidden by the robust limits on the effective
temperature, {\mk}, and distance.) Such a binary would produce 20--50\%
more than the 3$\sigma$ upper limit or 50--100\% more than the measured
bolometric luminosity for the region; thus, a correspondingly large
fraction must escape the {\UCHII} region.  A close binary might also
explain the broad photospheric lines seen in the spectrum.  Whether this
explanation is more acceptable than the one offered by Watson et~al.\
(1997) is left to the judgement of the reader.

If we are to accept the idea that the ionizing star in {\g} evolved
significantly as it accreted, we must provide an explanation for why
HD~93250 and O stars in M17 apparently have not (Hanson, Howarth,
\& Conti 1997). The time allowed for evolution to proceed during the
accretion phase is inversely proportional to the mean accretion rate.
It is possible, then, that accretion proceeded sufficiently quickly in
M17 and HD~93205 that evolutionary effects are not noticeable, but slowly
enough in {\g} that they are.  Models run at a number of accretion rates
at both solar and twice solar metallicity are needed to investigate this.

If the broad photospheric lines seen in our spectrum are the result of very
rapid rotation, the von Zeipel effect will act to reduce the effective
temperature of the equatorial regions.  This could reconcile our
luminosity and effective temperature with those of a single main-sequence
star.  However, we are doubtful that the effect can lower the temperature
by the several thousand degrees required.

\section{Nebular Models}

Simpson et al.\ (1995) and Afflerbach, Churchwell, \& Werner (1997)
have modeled the far infrared emission lines of {\g} and found best fits
with ionizing stars that have effective temperatures that are cooler
than our lower limit of 38500~K. Similarly, Faison et~al.\ (1997) have
recently modeled the dust re-emission of {\g} and find a best fit with
ionizing stars that have luminosities that are below our lower limit on
the luminosity of the region. Clearly, satisfying our constraints on the
stellar properties while correctly predicting the mid and far infrared
emission will be an excellent test of models of the nebular emission.
Success may require more complex models that pay attention to the geometry
of {\g}, in particular its lack of spherical symmetry and wide range
of ionization parameters.

In the meantime, we can use our limits to investigate the ionization in
the nebula.  The recombination rate of $1.35 \times 10^{49}~\rm s^{-1}$
at 5~kpc is a lower limit to the Lyman continuum photon flux of the
ionizing star, as dust can absorb ionizing photons and, possibly, ionizing
photons can escape from the nebula. We calculated this recombination
rate from the intrinsic {\Br} flux given by Watson et~al.\ (1997) and
the Case B nebular models of Hummer \& Storey (1987).  If we use a fit
to Figure 13 of Schaerer \& de~Koter (1997) to give $q_0(T)$ and $L =
4 \pi R^2 \sigma T^4$ to give the Lyman continuum photon flux $Q_0(L,T)
= 4 \pi R^2 q_0$, then we find that the Lyman continuum flux exceeds the
effective recombination rate by factors of about 1.5 (at O8) to about 4
(at O5). This suggests that either dust is a significant competitor for
ionizing photons or that a significant fraction of the ionizing photons
are escaping.

\section{Summary}

We have successfully obtained a {\K}-band classification
spectrum of the ionizing star of the {\UCHII} region {\g}. The lines
of {\HI} and {\HeI} are obliterated by nebular emission, but the
spectrum shows {\CIV} and {\NIII} emission and {\HeII} absorption.
Using the classification scheme of Hanson, Conti, \& Rieke (1996), we
are able to restrict the star to spectral classes of O5 to O8 which
correspond to effective temperatures of 46000~K to 38500~K.

When we combine our effective temperature limits with the absolute
magnitude determined by Watson et~al.\ (1997), we can place the star
on the H-R diagram using only the relatively well-understood
properties of the stellar photosphere and the ionized gas. Our limits
are more robust than previous ones, which relied on a mixture of
assumptions and models for the influence of the star on the
surrounding ISM.

One direct result of this work is to provide a test case for models of
the nebular line and dust emission; these  must now satisfy the
independent constraints on the effective temperature and luminosity of
the ionizing star. In this context, the recently obtained ISO SWS and
LWS spectra of {\g} will be a valuable resource. 

At moderate resolution ($\lambda/\Delta\lambda \sim 1000$), the strong
nebular emission in {\UCHII} regions obliterates the stellar {\HI} and
{\HeI} features leaving us with just the {\CIV}, {\NIII}, and {\HeII}
features in the {\K}-band. The spectral types that can be determined
from these features are quite rough: O3 to O4 ({\NIII} emission and
{\HeII} absorption but no {\CIV} emission), O5 to O8 (both {\NIII} and
{\CIV} emission and {\HeII} absorption), and O9 or later (none of
these). Observations at high resolution ($\lambda/\Delta\lambda \sim
10000$) may be able to resolve the broad stellar lines from the narrow
nebular lines, but will be challenging.

This is the first time that the effective temperature of the ionizing
star of a {\UCHII} region has been determined directly. This technique
may be feasible in other {\UCHII} regions, although nebular veiling
and higher extinction may be serious problems. This work is a further
demonstration that {\K}-band spectral classification, when combined with
near infrared imaging, is a uniquely powerful tool for the study of very
young massive stars.

\acknowledgements

We are grateful to George Rieke and Carol Heller for assistance at the
MMT, to Ed Churchwell, Joe Cassinelli, Jay Gallagher, and Don Garnett
for comments on this work, to Pepe Franco for forcing us to think about
long lived {\UCHII} regions, to Henny Lamers for suggesting we think
about the von~Zeipel effect, and to an anonymous referee for a number
of useful comments.

The observations reported in this paper were obtained at the
MMTO, a facility operated jointly by the University of Arizona and the
Smithsonian Institution. MMH is supported by NASA through a Hubble
Fellowship grant \#HF0107201.94A awarded by the Space Telescope
Science Institute, which is operated by the Association of
Universities for Research in Astronomy, Inc., for NASA under contract
NAS 5--26555.

%ms: \pagebreak
%ms: \pagestyle{empty}
%ms: \setcounter{topnumber}{20}
%ms: \setcounter{totalnumber}{20}
%ms: \section*{Figure Captions}

\pagebreak
\epsfxsize=\hsize
\epsfbox{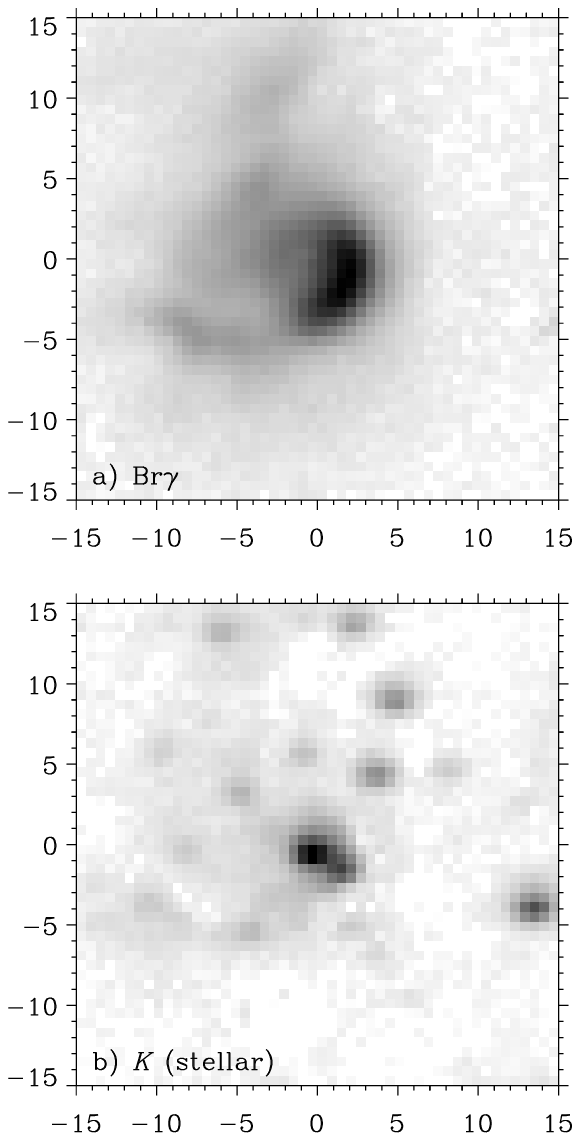}

\figcaption[image.eps]{(a) {\Br} and (b) {\K}-band images of the region
around {\g} from Watson et~al.\ (1997). The axes are marking arcsec. The
{\Br} image has been scaled and subtracted from the {\K} image to
suppress nebular emission. Note the arc of bright nebular emission,
the fan of fainter nebular emission extending to the north-west, and
the ionizing star at the center of the arc.\label{fig-image}}

\pagebreak
\epsfxsize=\hsize
\epsfbox{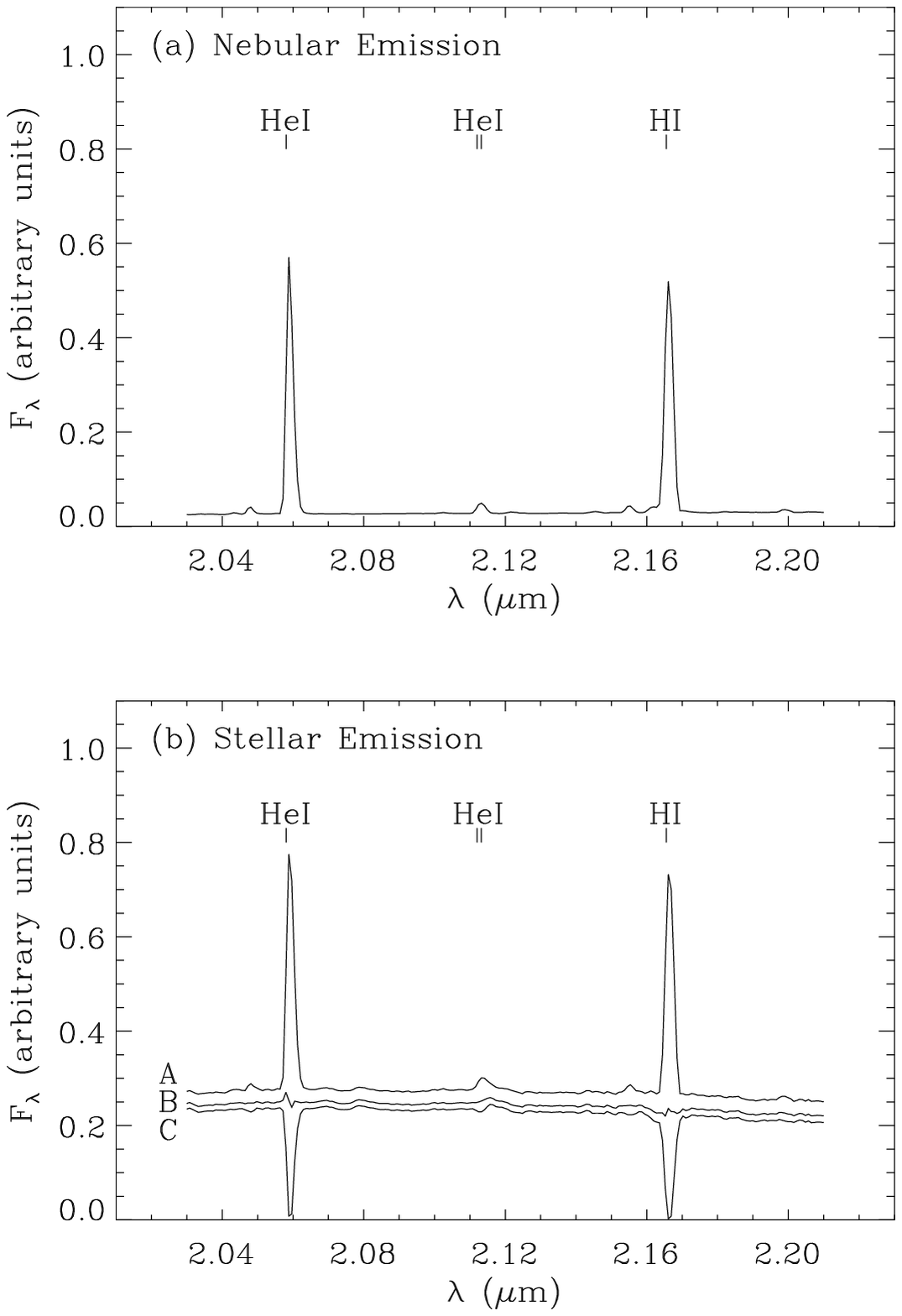}

\figcaption[flux.eps]{(a) The nebular arc spectrum. (b) The ionizing
star spectra. As described in Section~\ref{sec-data}, spectrum A is
not corrected for nebular emission, spectrum B is corrected to leave
the Br$\gamma$ and {\HeI} 2.0581{\micron} lines neither in emission or
absorption, and spectrum C is overcorrected.\label{fig-flux}}

\pagebreak
\epsfxsize=\hsize
\centerline{\epsfbox{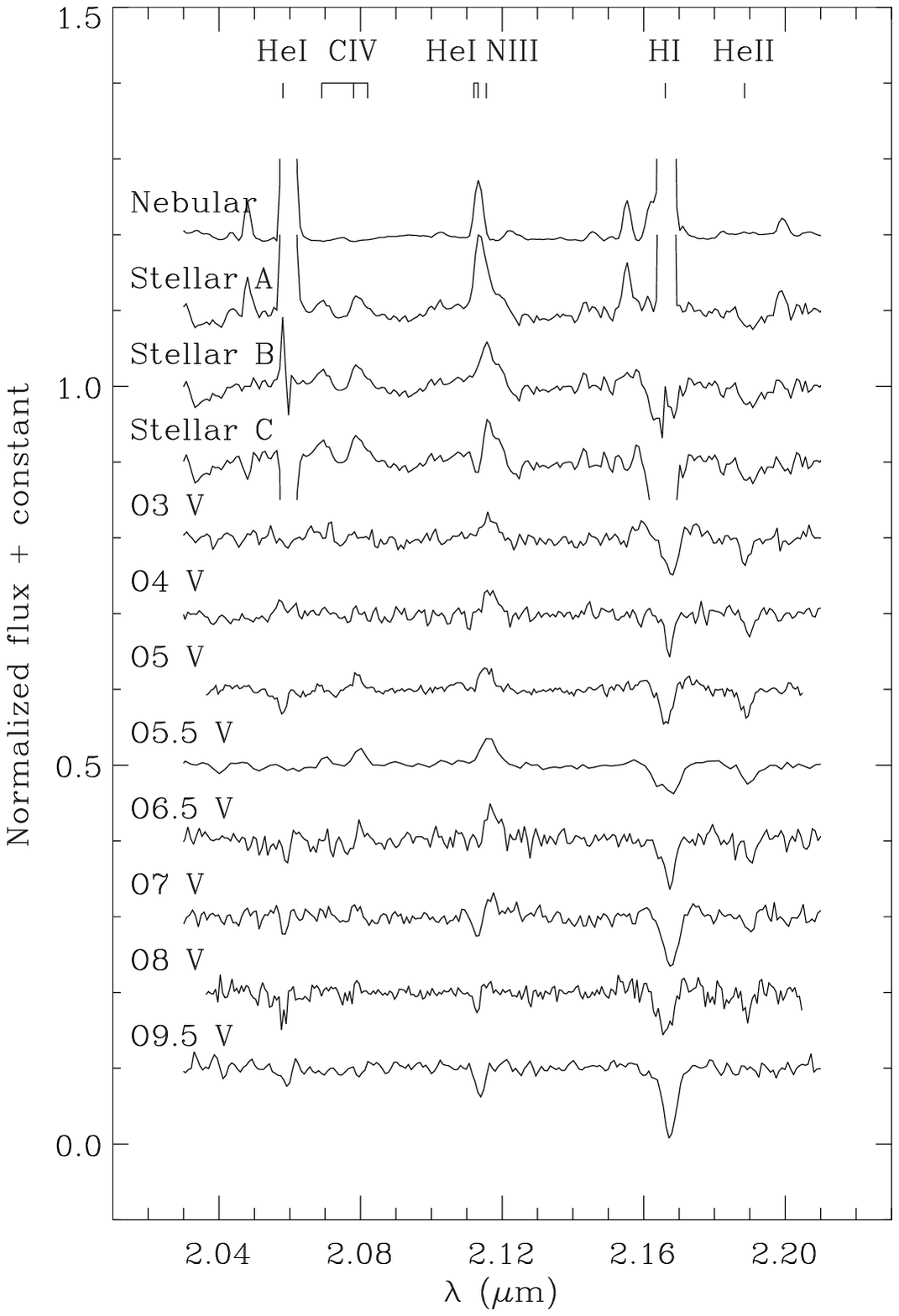}}

\figcaption[spec.eps]{Normalized spectra of the
nebula and ionizing star from this work and of comparison stars from
Hanson, Conti, \& Rieke (1996).\label{fig-spec}}

\pagebreak
\epsfxsize=\hsize
\centerline{\epsfbox{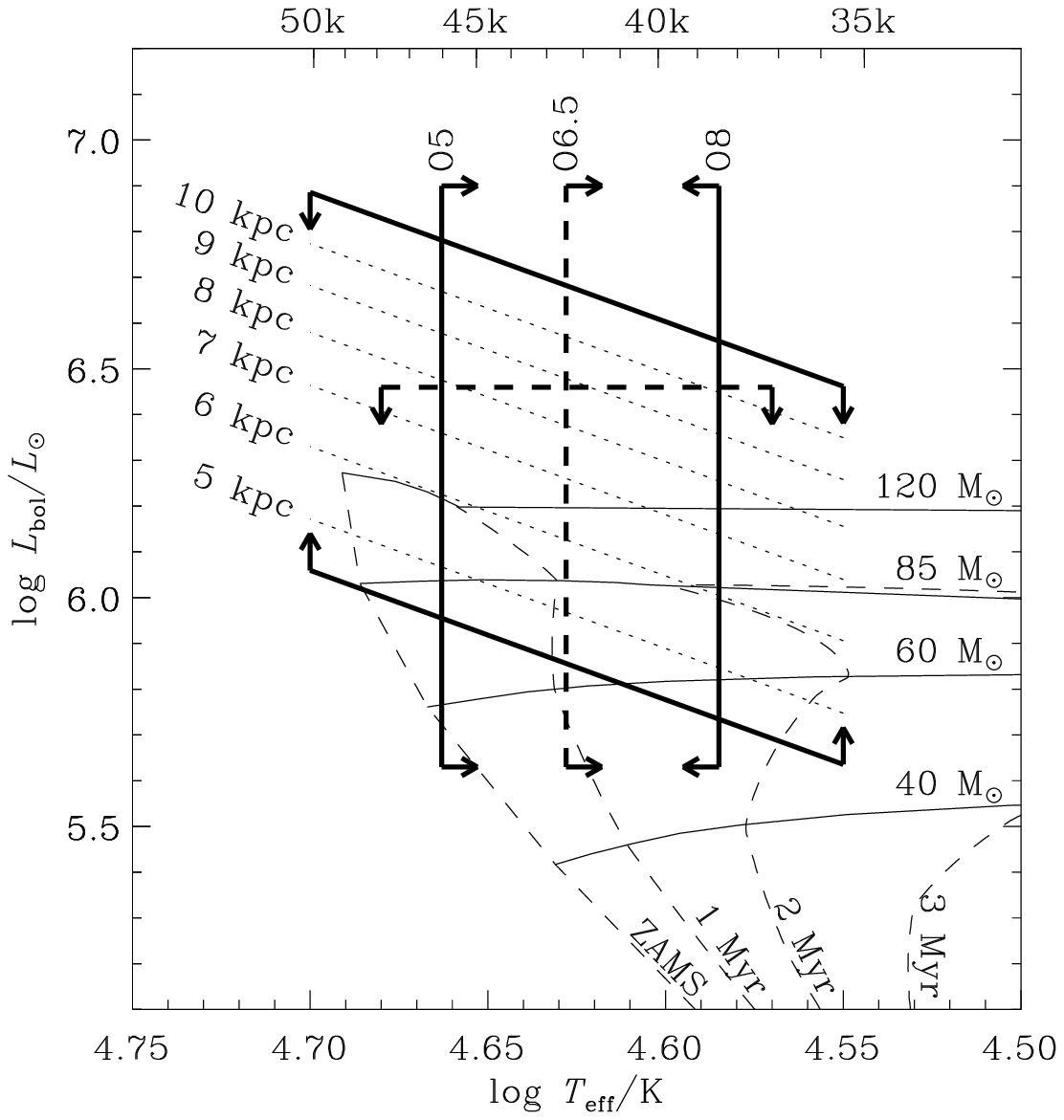}}

\figcaption[spec.eps]{A theoretical H-R diagram for the ionizing star.
The temperature limits derived from our spectral classification are
shown by thick solid vertical lines. The luminosity limits determined
by Watson et~al.\ (1997) from limits on the distance and {\mk} are
shown by sloping thick solid lines. The limits on the effective
temperature and luminosity derived by Watson et~al.\ (1997) from
limits on the distance, {\mk}, and {\mbol} are shown as thick dashed
lines. The thin dotted lines show the loci where the {\mk} is exactly
that measured at different distances. Also shown are tracks and
isochrones of the $Z = 2Z_\odot$ models of Meynet et al.\ (1994). The
thin solid lines are tracks for 120, 85, 60, and 40{\Msun} models and
the thin dashed lines are the ZAMS, 1, 2, and $3 \times 10^6~\rm yr$
isochrones.\label{fig-model}}

\end{document}